\documentclass[12pt,preprint]{aastex}

\usepackage{graphics}

\newcommand{\msun}{\mbox{$M_{\sun}$}}

\newcommand{\usec}{\mbox{$\mu$s}}

\shorttitle{Shapiro delay in the PSR~J1640+2224 system}
\shortauthors{O.~L\"ohmer et al.}

\begin{document}

\title{Shapiro delay in the PSR~J1640+2224 binary system}

\author{Oliver L\"ohmer}
\affil{Max-Planck-Institut f\"ur Radioastronomie, Auf dem H\"ugel 69,
       D-53121 Bonn, Germany}

\author{Wojciech Lewandowski}
\affil{Toru\'n Centre for Astronomy, Nicolaus Copernicus University, 
       Gagarina 11, 87-100 Toru\'n, Poland}

\author{Alex Wolszczan}
\affil{Department of Astronomy and Astrophysics, Pennsylvania State
           University, University Park, PA 16802, USA}
\affil{Toru\'n Centre for Astronomy, Nicolaus Copernicus University, 
       Gagarina 11, 87-100 Toru\'n, Poland}
\affil{Max-Planck-Institut f\"ur Radioastronomie, Auf dem H\"ugel 69,
       D-53121 Bonn, Germany}

\author{Richard Wielebinski}
\affil{Max-Planck-Institut f\"ur Radioastronomie, Auf dem H\"ugel 69,
       D-53121 Bonn, Germany}

\begin{abstract}
We present the results of precision timing observations of the binary
millisecond pulsar \object{PSR J1640+2224}. Combining the pulse
arrival time measurements made with the Effelsberg 100-m radio
telescope and the Arecibo 305-m radio telescope, we have 
extended the existing timing model of the pulsar to search for a
presence of the effect of a general-relativistic Shapiro delay 
in the data. At the currently attainable precision level, the
observed amplitude of the effect constrains the companion mass to
$m_2=0.15^{+0.08}_{-0.05}\, M_\sun$, which is consistent with the
estimates obtained from optical observations of the white dwarf
companion and with the mass range predicted by theories of binary
evolution. The measured shape of the Shapiro delay curve restricts the
range of possible orbital inclinations of the PSR J1640+2224 system to
$78^{\circ}\le i\le 88^{\circ}$. The pulsar offers excellent prospects
to significantly tighten these constraints in the near future.
\end{abstract}


\section{INTRODUCTION
\label{intro}}

Precision timing measurements of binary millisecond pulsars (Phinney
\& Kulkarni 1994\nocite{pk94}) with sufficiently high (near edge-on)
orbital inclinations make it possible to detect the effect of a
general-relativistic time delay of the pulsar signal in the
gravitational field of the companion star. For a pulsar in a circular
orbit this ``Shapiro delay'' (Shapiro 1964\nocite{sha64}) is given by
\begin{equation}\label{shap}
\Delta t = -2\, m_2\; T_{\sun}\, \ln[1-\sin i\, \sin(\Phi-\Phi_0)]\  ,
\end{equation}
where $\Phi$ is the orbital phase in radians, $\Phi_0$ is the phase of
the ascending node, and $T_{\sun}=(G\,M_{\sun}/c^3)$. In practice,
Shapiro delay is conveniently expressed in terms of two observables,
the ``range'' $r = m_2\,T_{\sun}$ and the ``shape'' $s = \sin i$
(Ryba \& Taylor 1991\nocite{rt91a}), the post-Keplerian orbital
parameters which allow a determination of the companion mass, $m_2$,
and the orbital inclination, $i$.

Because high inclination orbits are relatively rare, Shapiro delay has
been detected in only four pulsar-white dwarf (WD) binaries,
\object{PSR J1713+0747} (Camilo, Foster \& Wolszczan
1994\nocite{cfw94}), \object{PSR B1855+09} (Ryba \& Taylor 1991),
\object{PSR J0437$-$4715} (van Straten et al. 2001\nocite{vbb+01}),
\object{PSR J1909$-$3744} (Jacoby et al.\ 2003\nocite{jbv+03}), and
possibly in \object{PSR J0751+1807} (Nice, Splaver \& Stairs
2003\nocite{nss03}). Such detections offer very useful means to
measure masses of the companion stars and to calibrate the pulsar
spin-down models against the cooling models of white dwarfs. This is
accomplished by comparing the spin-down age of a pulsar obtained from
timing observations with the cooling age of a white dwarf estimated
from its mass, an optical measurement of its temperature and from an
appropriate cooling model (Kulkarni 1986\nocite{kul86}).  In
particular, such comparisons are important in assessing the
temperature modifying effect of hydrogen left over after the white
dwarf formation (van Kerkwijk et al.\ 2000\nocite{vbkk00}).

Shapiro delay has also been detected in two double neutron star
(NS--NS) systems, \object{PSR B1534+12} (e.g.\ Stairs et al.\
2002\nocite{stt+02}) and \object{PSR J0737$-$3039A} (Burgay et
al. 2003\nocite{bdp+03}; Lyne et al.\ 2004\nocite{lbk+04}). In these
cases, the measured parameters $r$ and $s$, together with the other
two strong gravity related post-Keplerian parameters, the periastron
advance $\dot\omega$, and the time dilation and gravitational redshift
$\gamma$, provide a ``clean'' test of general relativity and other
theories of gravity, in the sense that it does not mix the
relativistic strong-field and the radiative effects (Damour \& Taylor
1992\nocite{dt92}).

Neutron stars in pulsar--WD binaries are thought to undergo extended
periods of transfer of mass and angular momentum from their companions
(Phinney \& Kulkarni 1994). As a result, they are spun up to
millisecond periods and may end up having masses significantly larger
than the canonical value of $1.35\, M_{\sun}$ derived for stars in the
NS--NS systems (Thorsett \& Chakrabarty 1999\nocite{tc99}). The
existing mass measurements for PSR J0437$-$4715 ($m_1=1.58\pm
0.18\,M_{\sun}$, van Straten et al. 2001) and PSR B1855+09
($m_2=1.57^{+0.12}_{-0.11}\,M_{\sun}$) (Nice, Splaver \& Stairs 2003,
2004\nocite{nss04}) provide support for this idea. On the other hand,
the recent mass measurement of $1.3\pm 0.2\,M_{\sun}$ derived from
precision timing observations of PSR J1713+0747 (Splaver et al. 2004)
is in accord with the canonical value.  Undoubtedly, more data for
similar systems are needed to improve the existing statistics.

In this paper, we report new results of timing measurements of the PSR
J1640+2224 binary system. A tentative detection of the Shapiro delay
in the pulse arrival times from the pulsar allows us to set
preliminary constraints on the orbital inclination of the system and
the mass of the pulsar companion. In Sect.~\ref{obs} we describe the
timing observations made at Effelsberg and Arecibo, present the timing
analysis, and summarize the resulting best-fit timing model for PSR
J1640+2224. In Sect.~\ref{disc} we discuss the new findings and their
implications. In particular, we use simulated timing observations to
demonstrate the expected potential of a future sub-microsecond timing
of PSR J1640+2224 to verify the validity of the current best-fit model
and to improve the estimates of masses of the pulsar and its white
dwarf companion.

\section{OBSERVATIONS AND TIMING ANALYSIS \label{obs}}

We have conducted systematic, high-precision timing observations of
PSR~J1640+2224 with the 100-m Effelsberg radiotelescope of the
Max-Planck-Institut f\"ur Radioastronomie in Bonn, Germany, and the
305-m Arecibo radiotelescope of the National Astronomy and Ionosphere
Center in Puerto Rico, over a 7-year period from 1996 until 2003.

At Effelsberg, PSR J1640+2224 was observed approximately once a month
using a 1300$-$1700~MHz tunable HEMT receiver at a centre frequency of
1410~MHz. In order to monitor changes of the dispersion measure (DM)
we occasionally collected data at 860~MHz. As a backend, we used the
Effelsberg--Berkeley Pulsar Processor (EBPP), which corrects for the
dispersion smearing of the signal employing a coherent de-dispersion
technique (Hankins \& Rickett 1975\nocite{hr75}). In the total power
mode, the EBPP provided 32 channels for both senses of circular
polarization with a maximum total bandwith of 112~MHz, depending on
the DM and the observing frequency (Backer et al.\
1997\nocite{bdz+97}). For PSR~J1640+2224, total bandwidths of 54~MHz
and 27~MHz were available at 1410~MHz and 860~MHz, respectively. The
output signals of each channel were fed into the de-disperser boards
for coherent on-line de-dispersion and were synchronously folded at
the pulse period over a 7~min integration time.

At Arecibo, the timing observations of PSR J1640+2224 were made with
the dual-circular polarization receiving systems at 430 MHz, 1130~MHz
and 1410~MHz and the Penn State Pulsar Machine (PSPM). The PSPM pulsar
backend is a computer controlled processor with a $2\times 128\times
60$~kHz filterbank designed to conduct fast sampled pulsar searches
and precision timing measurements. Technical details of the backend
are given in Cadwell (1997\nocite{cad97a}). For our timing
observations, the two signals of opposite circular polarizations were
added together, smoothed with a 32 $\mu$s time constant, 4-bit
quantized, folded synchronously with the topocentric pulse period, and
stored for further processing. The pulse integration times were 3
min. at 430 MHz and 5 min. at both 1130 MHz and 1410 MHz.

Both Effelsberg and Arecibo data were time stamped using the
observatory hydrogen maser clocks and later synchronized to UTC(NIST)
using the signals from the Global Positioning System (GPS). In order
to calculate the pulse time-of-arrival (TOA), high signal-to-noise
template profiles of the pulse were constructed for each backend and
observing frequency and least-squares fitted to the observed profiles
in frequency domain (Taylor 1991\nocite{tay91}). A theory-independent
timing model for binary pulsars, devised by Damour \& Deruelle
(1986\nocite{dd86}), was least squares fitted to the combined TOAs,
weighted by their individual uncertainties, using the software package
{\sc tempo}\footnote{http://pulsar.princeton.edu/tempo} and the DE200
planetary ephemeris (Standish 1990\nocite{sta90}).

In the fitting procedure, the TOA segments obtained with the EBPP and
the PSPM were fitted for an unknown offset between the two data sets
resulting from different templates and TOA reference points in the
profiles. Using the full TOA set at all frequencies we determined the
DM of the pulsar. In the subsequent analysis, we fixed the best-fit
value for the DM and used only the 1410~MHz TOAs from Effelsberg and
both 1130 and 1410~MHz TOAs from Arecibo, as these high frequency data
were not significantly affected by the observed DM variations.  A
determination of an initial best-fit model for the PSR J1640+2224
timing data involved a set of 12 parameters including the astrometric,
and the rotational parameters of the pulsar and the orbital parameters
of the binary system. In order to achieve a uniform reduced $\chi^2=1$
for each data segment, we increased the TOA uncertainties by a
constant amount, approximately equal to the post-fit rms noise, by
adding it in quadrature to the actual TOA values.

\placefigure{\ref{fig:chi2}}

\begin{figure}
\includegraphics[angle=-90,width=14cm]{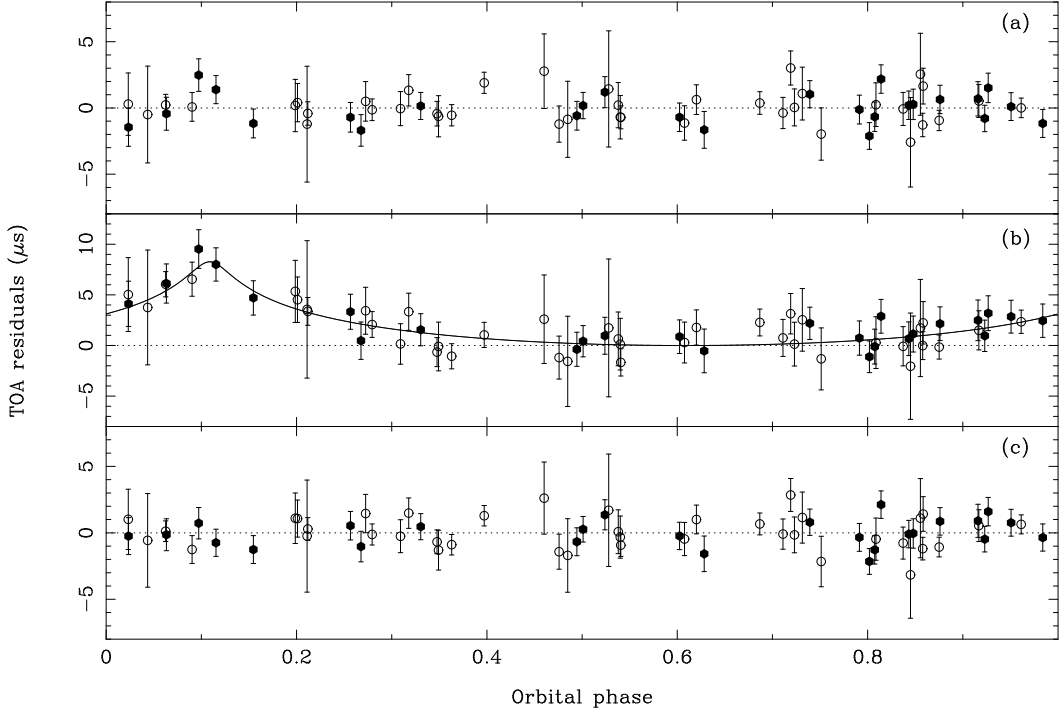}
\caption{ \label{fig:shapiro} Timing residuals for PSR J1640+2224
observed with the Arecibo telescope at 1130 MHz and 1410 MHz (filled
circles) and with the Effelsberg telescope at 1410 MHz (open circles),
as a function of orbital phase. (a) Post-fit residuals for the
best-fit model involving only the five Keplerian parameters.
(b) The effect of Shapiro delay on the
TOA residuals calculated with $\sin i$ and $m_2$ set to zero and all
other parameters fixed at their best-fit values.  The solid curve
represents the delay predicted by general relativity for the best-fit
Shapiro parameters. 
(c) Post-fit residuals for the best-fit model
including the $\sin i$ and $m_2$ parameters (see also
Table~\ref{tab:params}).
}
\end{figure}

In order to examine the timing data for a possible presence of
the Shapiro delay we employed a grid search procedure used by Ryba \&
Taylor (1991\nocite{rt91a}).  We searched the $m_2 - \cos i$ plane for
a global $\chi^2$ minimum, by fixing the Shapiro parameters $r$ and
$s$ at nodes of an appropriately defined 2D grid and repeatedly
fitting for all other parameters for each set of $(r,s)$ values.  As
displayed in Fig.~\ref{fig:chi2}, the grid search produces a
well-defined global $\Delta\chi^2$ minimum, equivalent to the best-fit
$\cos i=0.11^{+0.09}_{-0.07}$ ($i\sim 84^{+4}_{-6}$ degrees) and
$m_2=0.15^{+0.08}_{-0.05}\,M_{\sun}$ (1$\sigma$
uncertainties). Clearly, the inclusion of a Shapiro delay in the
timing model for PSR J1640+2224 leads to astrophysically plausible
estimates of both the companion mass and the inclination of the pulsar
orbit. In Fig.~\ref{fig:shapiro}, the timing residuals for the
combined Arecibo and Effelsberg observations are plotted as a function
of orbital phase. Because the observed Shapiro delay is weak and
both $r$ and $s$ are strongly covariant with other model parameters,
the effect is not detectable in residuals from the best fit involving
the Keplerian orbit alone, as seen in Fig.~\ref{fig:shapiro}a. On the
other hand, in Fig.~\ref{fig:shapiro}b, showing a Shapiro delay
signature extracted from the grid search with all other timing effects
removed (see also Ryba \& Taylor (1991) and Camilo, Foster \&
Wolszczan (1994)), the amplitude of the effect in the PSR J1640+2224
TOA residuals significantly exceeds the TOA uncertainties, as expected
from the result of the $\chi^2$ search displayed in
Fig.~\ref{fig:chi2}.

In principle, the observed signature could be induced by DM variations
over the pulsar orbit. If the pulsar's WD companion had an extended
envelope created by the pulsar wind and the high-energy photon flux,
as observed in eclipsing binary systems (e.g. Nice, Arzoumanian \&
Thorsett 2000\nocite{nat00}), the electron column density would
fluctuate periodically as a function of orbital phase. For highly
inclined orbits this would obviously cause periodic,
frequency-dependent TOA variations that could mimic the effect of
Shapiro delay. We have ruled out this possibility by verifying that
the effect has the same amplitude in the TOA measurements made at four
different frequencies.

In the case of binary pulsars with nearly circular, low-inclination
orbits, Shapiro delay becomes covariant with Roemer delay and cannot
be measured (Lange et al.\ 2001\nocite{lcw+01}). PSR J1640+2224 has
the most eccentric orbit among the pulsar--WD binaries (see
Table~\ref{tab:params} and Edwards and Bailes 2001\nocite{eb01b}) and
its inclination angle of $i=84\pm6\degr$ derived from our analysis
appears to be high enough to allow the inclusion of Shapiro
delay in the timing model (Fig.~\ref{fig:shapiro}b). In any case,
further observations of the pulsar with higher timing precision are
clearly necessary to fully assess a statistical significance of our
detection.

\placefigure{\ref{fig:shapiro}}

\begin{figure}
\includegraphics[angle=-90,width=12cm]{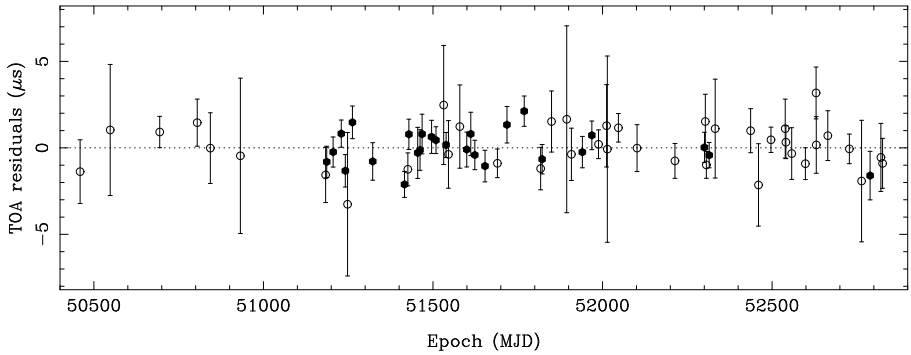}
\caption{ \label{fig:resid-epoch} Best-fit timing residuals for PSR
J1640+2224 as a function of observing epoch. See
Fig.~\ref{fig:shapiro} for further explanation.}
\end{figure}

\placefigure{\ref{fig:resid-epoch}}

The parameters of the best-fit timing model for PSR J1640+2224 are
listed in Table~\ref{tab:params} along with the ones for which only
upper limits could be determined. In this case, the upper limits were
obtained by allowing the parameters to vary, one at a time, in the
global fit.  Also included in the table are the most important
parameters derived from the final model. Finally, the behavior of the
post-fit timing residuals as a function of time, spanning a 7-year
period, is shown in Fig.~\ref{fig:resid-epoch}.  Evidently, the
best-fit model including the Shapiro delay leaves no additional
systematic effects above the current post-fit rms residual of 2.0
$\mu$s.

\placetable{\ref{tab:params}}

\section{DISCUSSION \label{disc}}

A new timing model for the binary millisecond pulsar PSR J1640+2224
discussed in this paper is entirely consistent with the previous
models published by Wolszczan et al.\ (2000)\nocite{wdk+00} and
Potapov et al.\ (2003)\nocite{pio+03}. In addition, owing to a higher
timing precision, the new model provides further reduction of the
parameter estimation errors and, above all, it includes
astrophysically sensible estimates of the Shapiro delay parameters.

\begin{figure}
\includegraphics[angle=-90,width=12cm]{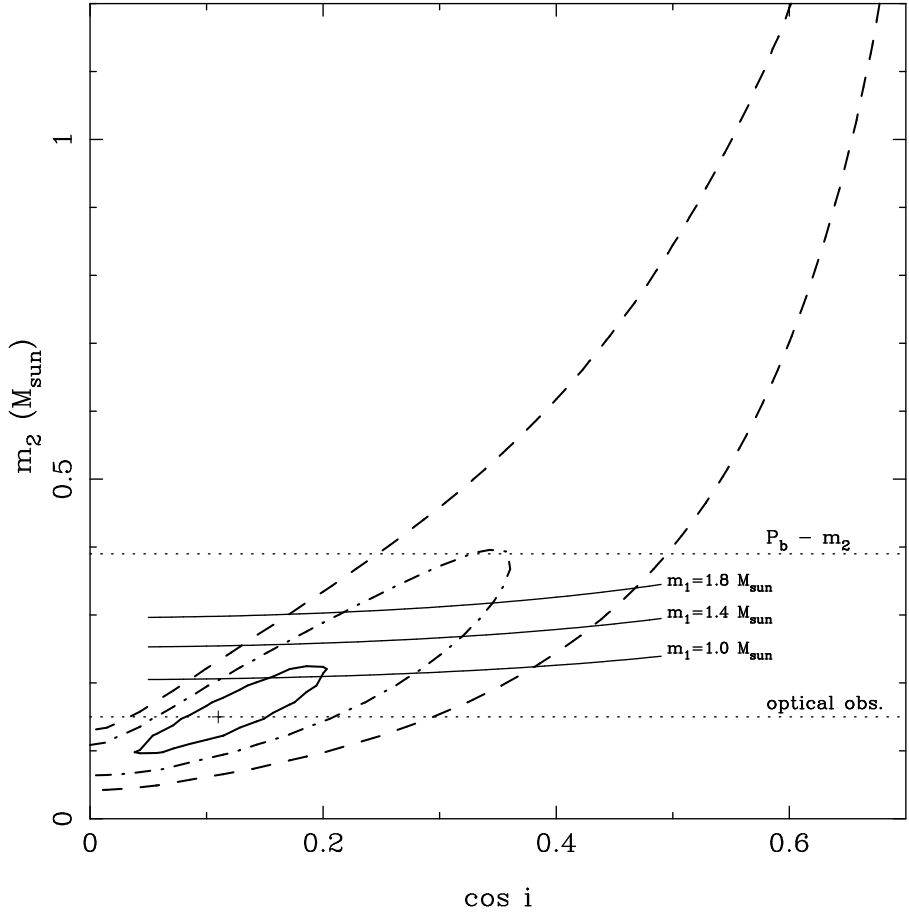}
\caption{ \label{fig:chi2} Constraints on $m_2$ and $\cos i$ in the
PSR J1640+2224 system from a $\chi^2$ search for the best-fit Shapiro
delay parameters. The global $\chi^2$ minimum is indicated by a cross.
The contours of $\Delta\chi^2 = 1.0$ (solid), $\Delta\chi^2 = 4.0$
(dashed-dotted) and $\Delta\chi^2 = 9.0$ (dashed) have extrema
respectively corresponding to $1\sigma$, $2\sigma$, and $3\sigma$
errors on the individual parameters $\cos i$ and $m_2$. Lines of
constant $m_1$ are indicated. Horizontal lines denote the $m_2$--range
bounded by a lower limit obtained from optical observations ($m_2\ge
0.15$\msun) and an upper limit allowed by the $P_b-m_2$ relationship
($m_2\le 0.39$\msun).}
\end{figure}

The binary companion to PSR J1640+2224 is a white dwarf with the
estimated cooling age and mass of 7$\pm$2 Gyr and 0.25$\pm 0.10\,
M_{\sun}$, respectively, as determined from the Hubble Space
Telescope observations (Lundgren et al.\ 1996\nocite{lfc96}).  A range
of masses predicted by the binary period--companion mass ($P_b-m_2$)
relationship based on the theory of low- and intermediate-mass binary
evolution is $0.35\le m_2\le 0.39\,M_{\sun}$ (Tauris \& Savonije
1999\nocite{ts99a}), and the minimum companion mass from the mass
function, $f(m_1,m_2)=(m_2 \sin i)^3(m_1+m_2)^{-2}=
(2\pi/P_b)^2x^3/T_{\sun}=0.0059\,M_{\sun}$ is $m_2=0.25$\,\msun\ for
a $m_1=1.4\,M_{\sun}$ neutron star. At the presently attainable level
of accuracy, the best-fit companion mass of
$m_2=0.15^{+0.08}_{-0.05}\,M_{\sun}$ derived from our data is
consistent with the above estimates, as illustrated in
Fig.~\ref{fig:chi2}.

\begin{figure}
\includegraphics[angle=-90,width=14cm]{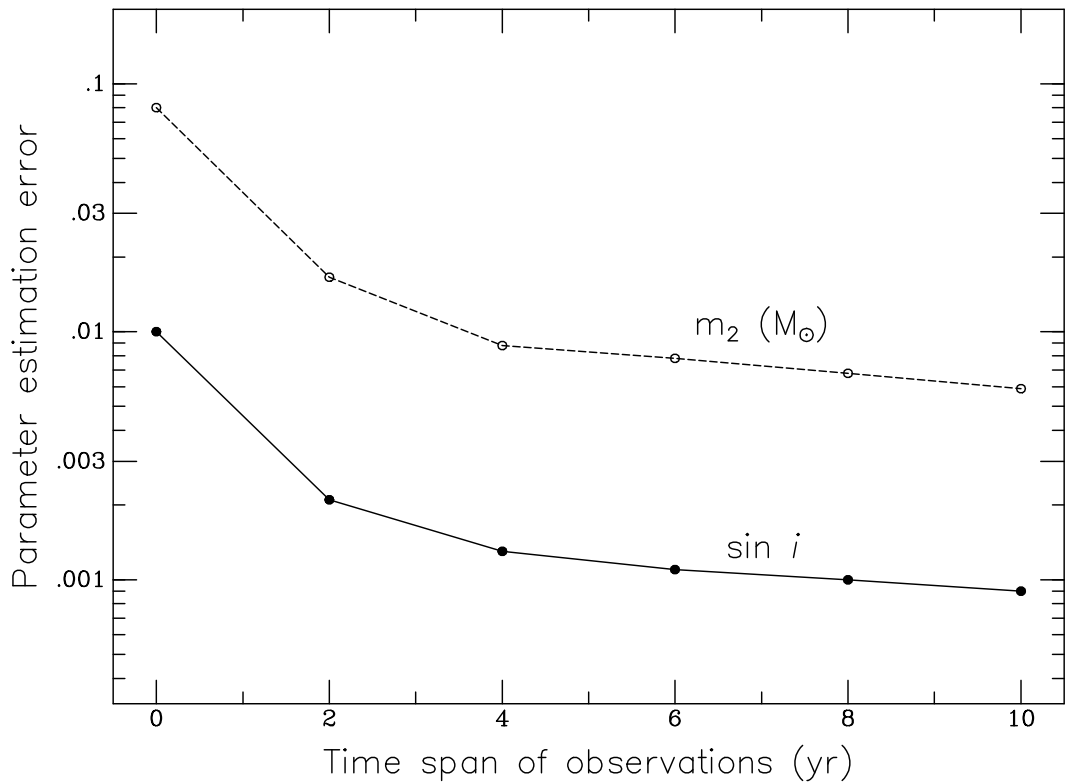}
\caption{ \label{fig:err} Estimation errors of the companion mass and
orbital inclination from the simulated TOA measurements of PSR
J1640+2224 with a 0.5 $\mu$s precision. The initial error values are
equal to those currently observed.}
\end{figure}

\placefigure{\ref{fig:err}}

Among the pulsar--WD binaries with a detectable Shapiro delay, only
PSR J0437$-$4715, PSR J1713+0747, and PSR B1855+09 have the values of
$m_2$ and $\sin i$ determined with the accuracy that is high enough to
make them usable in setting a tight constraint on the pulsar mass (van
Straten et al.\ 2001; Nice, Splaver \& Stairs 2003, 2004; Splaver et
al. 2004) and in investigating the details of evolution of the
pulsar's WD companion (van Kerkwijk et al. 2000).  We have examined a
future potential of the PSR J1640+2224 timing to become comparably
useful by generating artificial TOAs according to the model of
Table~\ref{tab:params} and analyzing the data over progressively
longer periods of time for several reasonable values of the timing
precision. Encouragingly, as shown in Fig.~\ref{fig:err}, in only four
years of monthly timing measurements with a 0.5 $\mu$s precision, the
estimation errors of $m_2$ and $\cos i$ approach the respective levels
of $0.01\,M_{\sun}$ and 0.001. Since the Arecibo timing measurements
using the PSPM and an 8 MHz receiver bandwidth are characterized by a
$\sim$1 $\mu$s long-term residual, it is quite conceivable that the
required $\le$0.5 $\mu$s precision can be achieved with a new
generation of broadband, 100 MHz bandwidth backends already available
at the telescope. Further observations at this level of precision will
quite conceivably allow a verification of the timing model presented
in this paper.  PSR J1640+2224 is also likely to become a valuable
member of a set of the most accurate pulsar clocks that can be timed
either individually, or as an array to detect a low-frequency
background of gravitational radiation (e.g. Thorsett \& Dewey
1996\nocite{td96}, Jaffe \& Backer 2003\nocite{jb03}).

\acknowledgments

We are very grateful to all staff at the Effelsberg and Arecibo
observatories for their help with the observations. We thank O.\
Doroshenko, A.\ Jessner and M.\ Kramer for their assistance in the
Effelsberg timing project and helpful discussions about the PSR
J1640+2224 binary system. A.W.'s research was supported by the
Alexander von Humboldt Foundation and the National Science Foundation
under Grant No. PHY99-07949. Arecibo Observatory is part of the
National Astronomy and Ionosphere Center, which is operated by Cornell
University under contract with the National Science Foundation.

\bibliographystyle{apj}

\clearpage

\clearpage

\begin{table*}
\begin{center}
\caption{Timing model for PSR~J1640+2224\label{tab:params}
}
\vspace{0.1cm}
\begin{tabular}{ll}
\hline\hline
\multicolumn{2}{c}{\it Measured parameters\,$^a$}\\
\hline
Right ascension, $\alpha$ (J2000) & $16\,\fh 40\,\fm 16\,\fs
742307(10)$ \\
Declination, $\delta$ (J2000)     & $22\degr 24\arcmin 08\,\farcs 9413(3)$\\
Proper motion, $\mu_{\alpha}$ (mas\,yr$^{-1}$)   & $1.66(12)$ \\
Proper motion, $\mu_{\delta}$   (mas\,yr$^{-1}$)   & $-11.3(2)$ \\
Pulse frequency, $\nu$ (s$^{-1}$)          & $316.123984313238(2)$\\
Pulse frequency derivative, $\dot\nu$ (10$^{-16}$ s$^{-2}$) & $-2.8257(9)$\\
Pulse period, $P$ (ms)                     & $3.16331581791380(2)$\\
Period derivative, $\dot P$ (10$^{-20}$ s\,s$^{-1}$) & $0.28276(9)$\\
Epoch (MJD)                              & $51700.0$ \\
Dispersion measure, DM (pc\,cm$^{-3}$) & $18.4260(8)$  \\
Orbital period, $P_{\rm b}$ (d)          & $175.46066194(7)$\\
Projected semi-major axis, $x$ (lt-s)    & $55.3297198(4)$    \\
Eccentricity, $e$                    & $0.000797262(14)$      \\
Epoch of periastron\,$^b$, $T_0$ (MJD)   & $51626.1785(5)$\\
Longitude of periastron\,$^b$, $\omega$ (deg)& $50.7308(10)$ \\
Shape of Shapiro delay\,$^c$, $s$ & $0.99^{+0.01}_{-0.01}$\\
Range of Shapiro delay\,$^c$, $r$ ($\mu$s)& $0.74^{+0.39}_{-0.25}$\\
\hline
\multicolumn{2}{c}{\it Measured upper limits\,$^d$}\\
\hline
Parallax, $\pi$ (mas) & $<3.7$\\
Pulse frequency second derivative, $|\ddot\nu|$ (s$^{-3}$)& $<4\times 10^{-27}$\\
DM derivative, $|d({\rm DM})/dt|$ (pc\,cm$^{-3}$\,yr$^{-1}$) & $<1.3\times 10^{-3}$\\
Orbital period derivative, $|\dot P_{\rm b}|$ (s\,s$^{-1}$)& $<3\times 10^{-10}$ \\
Derivative of projected semi-major axis, $|\dot x|$ (lt-s s$^{-1}$) &
$<1.7\times 10^{-14}$ \\
Periastron rate of change, $|\dot\omega|$ (deg yr$^{-1}$) & $<1.1\times 10^{-3}$ \\ 
\hline
\multicolumn{2}{c}{\it Derived parameters}\\
\hline
Galactic longitude, $l$  & $41\fdg 051$\\
Galactic latitude, $b$ & $38\fdg 271$\\
DM distance$^e$ (kpc)& $1.16$  \\
Composite proper motion, $\mu$ (mas\,yr$^{-1}$)   & 11.4(2)\\
Companion mass, $m_2$ (\msun)         &  $0.15^{+0.08}_{-0.05}$ \\
Orbital inclination angle, $i$ (deg)  &  $84^{+4}_{-6}$ \\
Mass function, $f_{\rm m}$ (\msun)     & $0.0059074304(2)$\\
Number of TOAs                        & 314\\
Timing RMS (\usec)           & 2.0        \\
\noalign{\smallskip}\hline
\end{tabular}
\end{center}
{\small $^a$ Figures in parentheses are $2\sigma$ uncertainties in the
last digits quoted (twice the formal {\sc tempo} errors).}\newline
{\small $^b$ $\omega$ and $T_0$ are highly covariant. Observers
should use $\omega=50.730834835740$ and $T_0=51626.178534099$}.\newline
{\small $^c$ Uncertainties are $1\sigma$ errors derived from the
$\chi^2$ analysis (see Sect.~2).}\newline
{\small $^d$ Upper limits represent 95\% C.L.}\newline
{\small $^e$ from the Cordes \& Lazio (2002\nocite{cl02a}) model of the
Galactic electron density distribution, with typical uncertainties of 
10\%}.\newline
\end{table*}

\end{document}